\documentclass[rnote,longauth,oldversion]{aa}

\usepackage{graphicx}
\usepackage{graphics}

\usepackage{longtable}
\usepackage{txfonts}

\usepackage{epsf,epsfig}
\usepackage[dvips]{color}
\usepackage{color}
\usepackage{amssymb,amsfonts}
\usepackage{latexsym}

\usepackage{tabularx}
\usepackage{multirow}

\usepackage{natbib}
\bibpunct{(}{)}{;}{a}{}{,}

\begin{document}

\newcommand{\ac}[1]{{\color{red} #1}}
\newcommand{\emma}[1]{{\color{blue} #1}}
\newcommand{\olaf}[1]{{\color{green} #1}}
\newcommand{\com}[2]{{\color{red} #1} {\color{green} #2}}
\newcommand{\fref}[1]{{Fig.~\ref{#1}}}
\newcommand{\tref}[1]{{Tab.~\ref{#1}}}

\hyphenation{msPSR}
\hyphenation{msPSRs}

\def\eg{e.g.}
\def\etal{et~al.}
\def\ie{{\em i.e. }}

\title{H.E.S.S. upper limit on the very high energy $\gamma$-ray emission from
the globular cluster 47~Tucanae}

\titlerunning{H.E.S.S. upper limit on the globular cluster 47~Tucanae}

\small{
\author{F. Aharonian\inst{1,13}
 \and A.G.~Akhperjanian \inst{2}
 \and G.~Anton \inst{16}
 \and U.~Barres de Almeida \inst{8} \thanks{supported by CAPES Foundation, Ministry of Education of Brazil}
 \and A.R.~Bazer-Bachi \inst{3}
 \and Y.~Becherini \inst{12}
 \and B.~Behera \inst{14}
 \and K.~Bernl\"ohr \inst{1,5}
 \and C.~Boisson \inst{6}
 \and A.~Bochow \inst{1}
% ...
  }
}

\offprints{clapson@mpi-hd.mpg.de}

\institute{
Max-Planck-Institut f\"ur Kernphysik, P.O. Box 103980, D 69029
Heidelberg, Germany
\and
 Yerevan Physics Institute, 2 Alikhanian Brothers St., 375036 Yerevan,
Armenia
\and
Centre d'Etude Spatiale des Rayonnements, CNRS/UPS, 9 av. du Colonel Roche, BP
4346, F-31029 Toulouse Cedex 4, France
\and
Universit\"at Hamburg, Institut f\"ur Experimentalphysik, Luruper Chaussee
149, D 22761 Hamburg, Germany
\and
Institut f\"ur Physik, Humboldt-Universit\"at zu Berlin, Newtonstr. 15,
D 12489 Berlin, Germany
\and
LUTH, Observatoire de Paris, CNRS, Universit\'e Paris Diderot, 5 Place Jules Janssen, 92190 Meudon, 
France
Obserwatorium Astronomiczne, Uniwersytet Ja
\and
IRFU/DSM/CEA, CE Saclay, F-91191
Gif-sur-Yvette, Cedex, France
\and
University of Durham, Department of Physics, South Road, Durham DH1 3LE,
U.K.
\and
Unit for Space Physics, North-West University, Potchefstroom 2520,
    South Africa
\and
Laboratoire Leprince-Ringuet, Ecole Polytechnique, CNRS/IN2P3,
 F-91128 Palaiseau, France
\and 
Laboratoire d'Annecy-le-Vieux de Physique des Particules, CNRS/IN2P3,
9 Chemin de Bellevue - BP 110 F-74941 Annecy-le-Vieux Cedex, France
\and
Astroparticule et Cosmologie (APC), CNRS, Universite Paris 7 Denis Diderot,
10, rue Alice Domon et Leonie Duquet, F-75205 Paris Cedex 13, France
\thanks{UMR 7164 (CNRS, Universit\'e Paris VII, CEA, Observatoire de Paris)}
\and
Dublin Institute for Advanced Studies, 5 Merrion Square, Dublin 2,
Ireland
\and
Landessternwarte, Universit\"at Heidelberg, K\"onigstuhl, D 69117 Heidelberg, Germany
\and
Laboratoire de Physique Th\'eorique et Astroparticules, CNRS/IN2P3,
Universit\'e Montpellier II, CC 70, Place Eug\`ene Bataillon, F-34095
Montpellier Cedex 5, France
\and
Universit\"at Erlangen-N\"urnberg, Physikalisches Institut, Erwin-Rommel-Str. 1,
D 91058 Erlangen, Germany
% ...
}

\date{\today}

\abstract{
Observations of the globular cluster 47~Tucanae (NGC 104), which
contains at least 23 millisecond pulsars, were performed with the
H.E.S.S. telescope system. The observations lead to an upper limit
of $F(E>800\,\mathrm{GeV}) < 6.7 \times 10^{-13}\,\mathrm{cm^{-2}\,s^{-1}}$
on the integral $\gamma$-ray photon flux from 47~Tucanae.
Considering millisecond pulsars as the unique
potential source of $\gamma$-rays in the globular cluster,
constraints based on emission models are derived:
on the magnetic field in the average pulsar nebula and on the conversion
efficiency of spin-down power to $\gamma$-ray photons or to relativistic leptons.
}

\maketitle

%%%%%%%%%%%%%%%%%%%%%%%%%%%%%%%%%%%%%%%%%
\section{Introduction}
\label{section:intro}

Millisecond pulsars (msPSRs) are usually categorized among the radio pulsar population by limits 
on their spin period ($P \leq 50\,\mathrm{ms}$) and, when available,
intrinsic spin-down rate ($\dot{P}_{int} \leq 10^{-18}\,\mathrm{s\,s^{-1}}$). 
They are old neutron stars, possibly re-accelerated by interactions with a companion, as first 
proposed in \citet{msPSRmodel}. Very high energy (VHE) emission from this type of object
has been predicted via various radiation mechanisms. For individual objects,
Inverse Compton (IC) or Curvature Radiation (CR) emission due to
the acceleration of leptons above the 
polar cap \citep{polarCap,ICflux} have been proposed. For binary
systems, an additional possibility would be the interaction between
pulsar wind driven outflows
and the stellar wind of the companion \citep[see for instance][]{PWNbinary}.
The spin-down power, typically lower than $10^{35}\,\mathrm{erg\,s^{-1}}$, entails
expected individual $\gamma$-ray fluxes well
below the detection threshold of current instruments.

However, groups of msPSRs have been identified in Galactic globular
clusters \citep[see \eg][]{firstGCpsrs}, allowing for larger fluxes from an ensemble of unresolved sources.
Out of the 185 pulsars with $P \leq 50\,\mathrm{ms}$ known in the year 
2008\footnote{http://www.atnf.csiro.au/research/pulsar/psrcat/ [v1.34]} \citep{atnf}, 
140 belong to globular clusters\footnote{http://www2.naic.edu/$\sim$pfreire/GCpsr.html [on 2008 August 7]}.
Globular clusters (GCs) are old high-density galactic structures, with
ages close to the age of the Galaxy itself \citep[see for instance][]{GCage}.
Their age indeed suggests evolved embedded stellar populations including compact (binary) objects,
which are considered potential progenitors to msPSRs, as discussed e.g. in \citet{LLRbinaries}.
The GCs Terzan~5, 47~Tucanae and M28, in this order, host the largest identified 
msPSR populations \citep{GCpulsars}.

47~Tucanae (NGC 104) is one of the largest Galactic GCs known to date,
with an estimated mass of $10^6\,\mathrm{M_\odot}$ and an
age of $11.2 \pm 1.1\,\mathrm{Gyr}$ \citep{GCage}. Optical
observations by the Hubble Space Telescope, described
in \citet{hst}, allowed precise measurements of its location,
centered at $\alpha_{2000}=0^h\,24^m\,05^s.67$ and
$\delta_{2000}=-72^{\circ}\,04'\,52''.62$ and placed it at a
distance of $4.0 \pm 0.35\,\mathrm{kpc}$. The surface brightness
distribution allows the estimation of 
a core radius of $r_0=20''.84 \pm 5''.05$, a half-mass radius $r_h \approx 2.6'$ and a
tidal radius $r_t \approx 0.6^{\circ}$, using the model of \citet{king66}.
In 47~Tucanae, 23 pulsars so far were revealed, with radio observations predominantly 
using the Parkes telescope \citep{parkes47tuc}, 
with periods in the range $2 - 8\,\mathrm{ms}$, averaging at $4\,\mathrm{ms}$,
all located within $1.2'$ of the centre of the GC.
Based on the unresolved $20\,\mathrm{cm}$ radio flux from the core of 47~Tucanae, \citet{radioObsmsPSR}
estimated that up to 30 pulsars could be radio-detected.
A study of the dispersion measure of the observed pulsar
period derivatives \citep{ionizedgas} provided an estimation of the
average msPSR intrinsic period derivative with
$\langle\dot{P} / P \rangle_{int} \approx 10^{-18}\,\mathrm{s^{-1}}$ and hence 
a surface dipole magnetic field $B_s \approx 2.6 \times 10^{8}\,\mathrm{G}$
and a spin-down power of $L_{sd} \approx 10^{34}\,\mathrm{erg\,s^{-1}}$.

At higher energies, \citet{chandrasurvey} reported, from Chandra X-ray observatory
data on 47~Tucanae, some 200 X-ray point sources, which belong
to several object classes including cataclysmic variables, low-mass
X-ray binaries (XRB), and the radio-detected msPSRs.
They derive, from a tentative identification of the unknown sources they detected,
an upper limit on the number of pulsars in the core of 47~Tucanae of about 60,
assuming individual fluxes similar to the X-ray detected ones.
Roughly two thirds of these msPSRs have a stellar companion ($M \le 0.2\,\mathrm{M_{\odot}}$).
The X-ray spectrum of a msPSR in a GC
can be described by a thermal component plus single power law,
with typical X-ray ($0.5$ -- $6\,\mathrm{keV}$) fluxes around $10^{31}\,\mathrm{erg\,s^{-1}}$ \citep{chandra19msp}.
A few msPSRs exhibit X-ray pulsations, although with pulsed fractions below
50\% for most of them \citep{variability}. 
The presence in 47 Tucanae of ``hidden'' msPSRs, detectable 
in hard X-rays but not in radio, has been excluded,
within the uncertainty of the model by \citet{hiddenmsPSR},
by the high-energy X-ray ($0.75$ to $30\,\mathrm{MeV}$) 
upper limits reported by COMPTEL \citep{comptelUL}. 
From EGRET observations, \citet{egretUL} produced a photon flux upper limit of
$5 \times 10^{-8}\,\mathrm{cm^{-2}\,s^{-1}}$ 
above $100\,\mathrm{MeV}$ at 95\% confidence level.
In the same energy band, a detection of 47Tuc was just announced with the release of
the FGST bright source
list\footnote{http://fermi.gsfc.nasa.gov/ssc/data/access/lat/bright\_src\_list/},
see \cite{BSLfermi}, slightly below the EGRET upper limit by \citet{egretUL}.
These are discussed in Section~\ref{section:discussion}.

In the TeV range, previous observations of globular clusters 
resulted in upper limits. A limit on the steady photon flux from M13 (5 msPSRs, $7\,\mathrm{kpc}$) 
was established by the Whipple Telescope \citep{M13whipple} at
$1.08\,\times\,\mathrm{10^{-11}\,cm^{-2}\,s^{-1}}$ above $500\,\mathrm{GeV}$. 
47~Tucanae was observed by the Durham~Mark~III telescopes, with a resulting upper limit
on the photon flux in pulsed emission from selected pulsars of 
$4.4\,\times\,10^{-11}\,\mathrm{cm^{-2}\,s^{-1}}$, 
above a threshold of $450\,\mathrm{GeV}$ \citep{47TucDurham}.
Periodic VHE emission from an XRB in 47~Tucanae above $5\,\mathrm{TeV}$
was reported once, by \citet{pulsedXRB} during a remarkably high X-ray flux
episode \citep{pulsed47tuc}. Such event has not been reported since then in 47~Tucanae.

The large number of identified msPSRs in 47~Tucanae and the
compactness of the msPSR population at relatively close 
distance motivated H.E.S.S. observations of this GC, to investigate the predicted VHE emission 
from this class of objects.
The results of these observations are presented in Section~\ref{section:observation}. 
Given the unknowns regarding VHE-emitting XRB, the interpretation
given in Section~\ref{section:discussion} centers
on a collective signature from the msPSR population at TeV energies. 

%%%%%%%%%%%%%%%%%%%%%%%%%%%%%%%%%%%%%%%%% 
\section{Observations and Analysis}
\label{section:observation}

H.E.S.S. is an array of four Imaging Atmospheric Cherenkov Telescopes,
located in the Khomas Highland of Namibia. 
Stereoscopic analysis methods allow efficient background (cosmic ray) rejection and accurate 
energy and arrival direction reconstruction for $\gamma$-rays in the
range $100\,\mathrm{GeV} - 100\,\mathrm{TeV}$.
For point-like sources, the system has a detection sensitivity of 1\% of the flux level of the Crab
Nebula above $1\,\mathrm{TeV}$ with a significance of $5\,\sigma$ in 25 hours of observation.
A thorough discussion of the H.E.S.S. standard analysis and performance of
the instrument can be found in \citet{crabhess}.

A total of 13 hours of 4-telescope data have been taken by
H.E.S.S. between October and November 2005 on 47~Tucanae 
(excluding data taken during bad weather or affected by hardware irregularities). 
The target was observed with an average zenith angle of $50\,^{\circ}$ and mean target
offset of $1\,^{\circ}$ from the centre of the field of view.
Applying the H.E.S.S. analysis ``standard'' cuts for point-like sources \citep[see][]{crabhess},
the energy threshold is about $800\,\mathrm{GeV}$ and 
the point-spread function above $1\,\mathrm{TeV}$ is $0.11^{\circ}$,
too large to resolve the core of 47~Tucanae.
Tighter cuts would slightly improve the sensitivity and angular resolution but also 
increase the energy threshold, further reducing the chances of a detection according to the
models (see Section~\ref{section:discussion}).
Several methods for $\gamma$-ray reconstruction \citep[the Hillas parameters method and a 
semi-analytical approach described in][]{modeleReco} and background estimation
 \citep[the ``ring'' and ``reflected'' algorithms discussed
in][]{crabhess} were used, with consistent results.

\begin{figure}
\centering
\includegraphics[width=9cm]{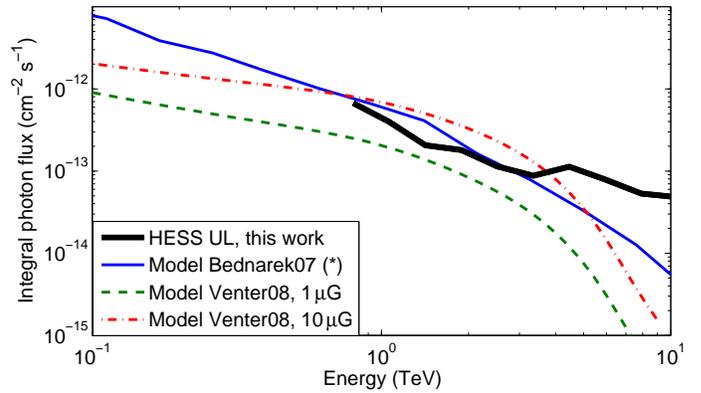}
\caption{Upper limit integral flux curve derived from the H.E.S.S. observations of 47~Tucanae 
(assuming a photon index of $\alpha = 2$), for ``standard'' cuts, at the 99\% confidence level. 
Predicted fluxes for 100 msPSRs were added for comparison, rescaled for a distance of $4\,\mathrm{kpc}$. 
(*) Curve adapted from \citet{bednarek07}, for $\epsilon_e = 0.01$, $E_{min} = 100\,\mathrm{GeV}$ and $\alpha = 2$,
rescaled to $L_{sd} = 10^{34}\,\mathrm{erg\,s^{-1}}$ (see Section~\ref{section:discussion} for details).}
\label{figure:intULmodels}
\end{figure}

We find no significant $\gamma$-ray event excess over the estimated background 
from the direction of 47~Tucanae.
With standard cuts and using the ``reflected''
background estimation method, the significance of the excess in the
$0.11\,^{\circ}$ radius integration area is $0.7\,\sigma$.
This allows us to set an upper limit on the flux from the target region. 
We determined upper limits according to \citet{FC-UL} with a 99\% confidence level,
assuming a point-like source and a power law photon flux energy spectrum of index $\alpha = 2$.  
The integral flux upper limit discussed here and shown in~\fref{figure:intULmodels}
was derived using the standard Hillas analysis,
consistent within 20\% with cross-check analyses.
Increasing the photon index to $\alpha = 3$ does not modify the
result by more than 20\%. The upper
limit on the integral photon flux in the H.E.S.S. energy range for this
data set ($800\,\mathrm{GeV}$ -- $48.6\,\mathrm{TeV}$, from the energy range of the 
collected events) is
$6.7 \times 10^{-13}\,\mathrm{cm^{-2}\,s^{-1}}$ or $\sim 2\,\%$ of the Crab flux.
This translates into a limit on the energy flux in the same energy range
of $6.8 \times 10^{33}\,\mathrm{erg\,s^{-1}}$ when placing 47~Tucanae at $4\,\mathrm{kpc}$ distance.
We also investigated an extended region ($0.2^{\circ}$ radius), without finding a significant excess.
We do not discuss the extended case further due to the compact distribution of the msPSRs 
in 47~Tucanae and the generally weaker limits derived for extended regions.

%%%%%%%%%%%%%%%%%%%%%%%%%%%%%%%%%%%%%%%%%
\section{Discussion}
\label{section:discussion}

The H.E.S.S. upper limit on the $\gamma$-ray flux emitted by 47~Tucanae 
can be confronted with scenarios of VHE
$\gamma$-ray emission by msPSRs involving accelerated leptons in
progressively larger regions: 
close to the pulsar, inside the pulsar wind nebula (PWN),
at the boundary of the eventual PWN,
or further away in the GC where pulsar winds may interact.
The comparison to PWNs detected in the VHE range is also discussed.
We only consider here average properties of the msPSRs in 47~Tucanae,
as summarized in Section~\ref{section:intro}, for populations of 23 (detected) or 100 sources.
While observational results favor smaller numbers,
results from dynamical models of GCs, \eg~from \citet{GCmodel}, suggest possibly larger populations.
Unless stated otherwise, the following constraints scale linearly with the number of pulsars.

The production of $\gamma$-rays in the pulsar magnetosphere has been proposed 
in (at least) two different general scenarios, which consider
different production sites: the ``outer gap'' or the ``polar cap''.
In the ``outer gap'' model \citep[see \eg][]{outerGap}, low values of the surface magnetic
field (estimated from the spin-down rate) and pulsar period, which define the conditions 
near the light cylinder, are believed to generally prevent VHE emission from msPSRs. 
Although the ``polar cap'' model \citep[discussed for instance in][]{polarCap} does not have such restriction 
on the conditions for VHE emission, both classes of model predict the flux to
drop off sharply between $1$ and $100\,\mathrm{GeV}$, as discussed
for a single msPSR in \citet{PSRmagnetosphere} and in \citet{GCdiffuse} for a large population. 
The upper limit by EGRET \citep{egretUL} does constrain some of these models. 
Pulsed emission is also predicted, e.g. in \citet{venter08} for 47~Tucanae, 
to drop before $100\,\mathrm{GeV}$, below the limit by \citet{47TucDurham}.
The Fermi detection will undoubtedly renew the discussion on these processes,
but interpretation in the $20\,\mathrm{MeV}$ -- $300\,\mathrm{GeV}$ band 
will be challenging, between potentially pulsed emission from one
or more msPSR, confused or unresolved sources, 
and the overall steady emission component, which could be tied to 
scenarios also valid at energies above our quoted threshold.

\begin{figure}
\centering
\includegraphics[width=9cm]{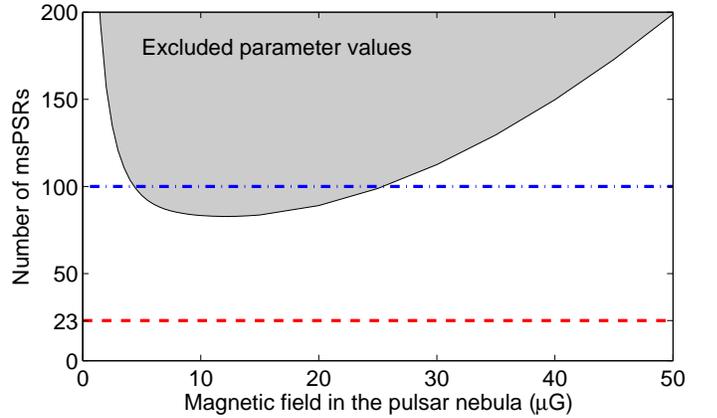}
\caption{Upper limit on the number of msPSRs in 47~Tucanae for a given average
magnetic field in the pulsar nebula, using the model by \citet{venter09} and the H.E.S.S. 
flux upper limit. The dashed lines indicate the number of observed msPSRs (23) and the 100
msPSRs hypothesis discussed here. 
\label{figure:Blimit}
}
\end{figure}

The IC component, produced either in the magnetosphere or further away from the compact object, 
does extend to the energies considered here, but in most cases with only very low fluxes \citep{ICflux}.
Still, when considering populations of sources, as done in \citet{venter08,venter09},
IC emission from the pulsar nebulae might reach observable levels.
Their Monte Carlo simulations of msPSR populations, 
accounting for the observed range of parameters ($P$, $\dot{P}_{int}$, viewing geometry),
predict the cumulative flux from 100 msPSRs, as illustrated in~\fref{figure:intULmodels}.
The efficiency of the IC emission in the
pulsar nebula increases with the strength of the nebular magnetic
field $B$ (in relation with the increased confinement time) until losses by
synchrotron radiation become dominant.
For a given value of $B$, the H.E.S.S. upper limit can be normalized by the predicted
flux per pulsar to obtain the maximum allowed number of msPSRs, as done in~\fref{figure:Blimit}.
Large msPSR populations are thus excluded, down to 80 objects for $B = 12\,\mathrm{\mu G}$.
In this model, the prediction falls short of providing constraints for only 23 msPSRs.
Using the limit on the magnetic field strength in the nebula -- post-shock -- of the
millisecond PSR~J0437-4715 by \citet{msp-Bsurf}
of $B < 18\,\mathrm{\mu G}$ and possibly lower (see the discussion in that reference),
and assuming similar properties for 100 msPSRs,
this would suggest $B \le 5\,\mathrm{\mu G}$ in the average pulsar nebula.

Another scenario for producing VHE $\gamma$-ray emission relies on particle acceleration 
at the shock discontinuity of a PWN.
Thorough discussions on pulsar winds can be found in \citet{PWNreview}.
In X-rays, pulsar wind emission from msPSRs has been observed, 
in the so-called ``black widow'' discussed in \citet{widowPWN}, 
with luminosities similar to those of canonical pulsars, but
not in a GC.
In this object, as well as for the ``Mouse'' pulsar \citep{mousePWN},
there are indications of interaction with the interstellar medium,
suggesting a bow shock geometry primarily driven by the proper motion
of the pulsar rather than by its accelerated particles. However, \citet{PWNinterDist} 
established that in a GC such bow shock emission would be hampered 
by the geometry and stellar density.
VHE $\gamma$-ray emission from several PWNs has already been detected and
\citet{VelaModel} suggested that msPSRs host the same leptonic emission processes as 
young pulsars like Vela~X ($290\,\mathrm{pc}$, $L_{sd} \approx 10^{36}\,\mathrm{erg\,s^{-1}}$).
Without assuming a particular emission process 
\citep[see][for a hadronic VHE emission model for the Vela~X PWN]{hadronicVelaX}, 
we derive the flux expected if similar objects were located in 47~Tucanae.
The VHE detection of the Vela~X PWN \citep{VelaX} gives an integral photon flux 
$F(E>800\,\mathrm{GeV}) \approx 1.5 \times 10^{-11}\,\mathrm{cm^{-2}\,s^{-1}}$. 
Scaling for the distance and spin-down power of the pulsar associated with the Vela~X 
nebula to the pulsars in 47~Tucanae amounts to a factor $5.3 \times 10^{-5}$.
We cannot constrain this model, as 840 ``Vela-like'' msPSRs would 
be required to reach our flux upper limit.
From pulsar properties and measured fluxes, it is usual to estimate the fraction of the spin-down power 
converted to $\gamma$-rays, $\epsilon_{sd}$, as compiled recently in \citet{PWN-VHE}
for the VHE-detected PWNs in the $1$ -- $10\,\mathrm{TeV}$ energy band.
For 47~Tucanae, this fraction is limited by $N_p \times \epsilon_{sd}^{1-10} \leq 0.7$.
Any msPSR population with $N_p \ge 23$ gives $\epsilon_{sd}^{1-10}$ in the broad range 
of detected PWNs ($8 \times 10^{-5}$ to $0.05$).
From this point of view, the msPSRs in 47~Tucanae cannot be distinguished from 
the much younger and more energetic pulsars detected through the VHE emission of their PWN.
Detailed studies of the specificities of each PWN might clarify the picture.
\begin{table}[t]
\renewcommand{\arraystretch}{1.3}
\centering
\begin{tabular}[center]{|*{2}{p{1.1cm}|}*{2}{p{0.7cm}|}*{1}{c}|}
\hline
\multicolumn{4}{|c|}{Model (\ddag) [$E_{min}\,(\mathrm{GeV})$, $\alpha$]} & Measured $\langle L_{sd} \rangle$ \\
100, 2.1 & 100, 3.0 & 1, 2.1 & 1, 3.0  & \\
\hline
\multicolumn{4}{|c|}{$\epsilon_e$} & $\epsilon_{sd}^{1-10}$ \\
\hline
 0.003 & 0.01 & 0.01 & 0.6 & 0.007 \\
\hline
\end{tabular}
\caption[47Tuc]{Upper limits on conversion efficiencies from spin-down power.
\label{table:constraint}
}
\end{table}

Nonetheless, a scenario by \citet{bednarek07} proposes that the energy of primary particles 
for the IC process increases through the interaction of the leptonic pulsar winds 
inside the GC. No observational evidence for such wind-wind interaction has yet been found.
They predict appreciable VHE $\gamma$-ray fluxes for a population of
100 msPSRs when the power emitted by each pulsar is fixed at 
$1.2 \times 10^{35}\,\mathrm{erg\,s^{-1}}$.
The distribution in energy of the leptons produced by a pulsar is assumed to follow a power law
of index $\alpha$, above a minimum energy $E_{min}$.
In most cases, the predicted flux in the H.E.S.S. energy range for 47~Tucanae should be above the 
detection threshold of the instrument.
According to this model, a non-detection translates in a limit on 
$N_P \times \epsilon_e$, the number of pulsars times the conversion efficiency
from the pulsar spin-down power into relativistic electron-positron pairs 
(and not $\epsilon_{sd}$, from spin-down to photons).
The available HESS data on 47~Tucanae do not allow to reach the reference sensitivity
used by \citet{bednarek07},
estimated (for 50 h of observation at $20\,^{\circ}$ zenith angle and $0.5\,^{\circ}$ offset) 
as a photon flux of about
$2.0 \times 10^{-13}\,\mathrm{cm^{-2}\,s^{-1}}$ above $800\,\mathrm{GeV}$, 
a factor $f_{sens} \approx 3.35$ lower than the result presented here. 
Besides, the values assumed in \citet{bednarek07} for the distance to 47~Tucanae 
($4.5\,\mathrm{kpc}$) and the individual spin-down power ($1.2 \times 10^{35}\,\mathrm{erg\,s^{-1}}$) 
may be too large. Overall, a factor 
$f_{sens} \times (L_{sd}^{Bednarek07} / L_{sd}^{data}) \times (d^{data}/d^{Bednarek07})^2 \approx 31.8$
must be applied when comparing their model predictions to the presented H.E.S.S. upper limit.
Since their original limit is on $N_P \times \epsilon_e$, a linear
rescaling can be applied when changing the number of pulsars.
Rescaled conversion efficiencies, derived from the H.E.S.S. upper limit
above $800\,\mathrm{GeV}$ assuming 100 msPSRs, are given in~\tref{table:constraint}
for their model (noted \ddag) and for the conversion from spin-down power to VHE
emission ($\epsilon_{sd}^{1-10}$) discussed above.
The comparison depends on the injection spectrum of the leptons produced by the pulsars. 
All the proposed scenarios are constrained ($\epsilon_e < 1$) in the 100 msPSRs case,
with most limits on the efficiency clearly below the estimated $\epsilon_e \approx 0.1$ 
for the Crab nebula \citep{bednarek07}, even when assuming only 23 msPSRs.
The exception is the scenario where most of the leptons are produced 
with low energy ($E_{min} = 1\,\mathrm{GeV}$ and $\alpha = 3$):
the constraints weaken to $\epsilon_e \le 0.6$ for 100 msPSR
and $\epsilon_e \ge 1$ (no constraint at all) for 23 msPRSs.

%%%%%%%%%%%%%%%%%%%%%%%%%%%%%%%%%%%%%%%%%
\section{Conclusions}

The upper limit of the VHE $\gamma$-ray photon flux  
obtained from H.E.S.S. observations of 47~Tucanae, 
$F(E > 800\,\mathrm{GeV}) < 6.7 \times 10^{-13}\,\mathrm{cm^{-2}\,s^{-1}}$,
is at present the second limit for a GC with a sizable population of 
msPSRs. Given the size of this population, it is the most 
constraining upper limit on the flux from an ensemble of msPSRs so far derived.

Comparing this result to emission models, we considered msPSRs as the only 
potential $\gamma$-ray sources in the GC.
Owing to the high energy threshold of these observations, emission models 
for the pulsar polar region, generally predicting low fluxes at these energies, cannot be
constrained, except when assuming msPSR populations much larger than considered 
here (23 -- 100).
These numbers, according to \cite{venter09}, may however be sufficient for 
the total IC emission to reach flux levels where the number of pulsars can be limited,
depending on the strength of the magnetic field in the pulsar nebula, down to
$B \le 5\,\mathrm{\mu G}$ in the average pulsar nebula for 100 msPSRs.
The limit on the conversion efficiency from spin-down power to VHE flux
(see~\tref{table:constraint}) is compatible with the results available for VHE-detected PWNs.
Collective IC emission as proposed
by \citet{bednarek07} cannot be more efficient than in the Crab nebula for most of their sets of parameters.
Complementary constraints at lower energy 
should follow the detection of 47~Tucanae by the Fermi Large Area
Telescope, but given the possible complexity of the emission in the GeV range,
the connection to the VHE band cannot be assessed here.

\begin{acknowledgements}
The support of the Namibian authorities and of the University of Namibia
in facilitating the construction and operation of H.E.S.S. is gratefully
acknowledged, as is the support by the German Ministry for Education and
Research (BMBF), the Max Planck Society, the French Ministry for Research,
the CNRS-IN2P3 and the Astroparticle Interdisciplinary Programme of the
CNRS, the U.K. Science and Technology Facilities Council (STFC),
the IPNP of the Charles University, the Polish Ministry of Science and 
Higher Education, the South African Department of
Science and Technology and National Research Foundation, and by the
University of Namibia. We appreciate the excellent work of the technical
support staff in Berlin, Durham, Hamburg, Heidelberg, Palaiseau, Paris,
Saclay, and in Namibia in the construction and operation of the equipment.
\end{acknowledgements}

\bibliographystyle{aa}

\end{document}